\documentclass[twocolumn]{aastex63}
\usepackage{graphicx}
\usepackage{xcolor}

\usepackage{CJKutf8}
\newcommand{\cntext}[1]{\begin{CJK}{UTF8}{gbsn}#1\end{CJK}\kern-1ex}

\shorttitle{Hot Plasma Flows and Loop-top Oscillations in a Solar Flare} 
\shortauthors{Reeves et al.} 

\begin{document}

\title{Hot Plasma Flows and Oscillations in the Loop-top Region During the September 10 2017 X8.2 Solar Flare}
\correspondingauthor{Katharine K. Reeves}
\email{kreeves@cfa.harvard.edu}

\author[0000-0002-6903-6832]{Katharine K. Reeves}
\affiliation{Harvard-Smithsonian Center for Astrophysics, 60 Garden St., Cambridge, MA 02138, USA}

\author[0000-0002-4980-7126]{Vanessa Polito}
\affiliation{Bay Area Environmental Research Institute, NASA Research Park,  Moffett Field, CA 94035-0001, USA}
\affiliation{Lockheed Martin Solar and Astrophysics Laboratory, Building 252, 3251 Hanover Street, Palo Alto, CA 94304, USA}

\author[0000-0002-0660-3350]{Bin Chen (\cntext{陈彬})}
\affiliation{Center for Solar-Terrestrial Research, New Jersey Institute of Technology, 323 M L King Jr. Blvd., Newark, NJ 07102-1982, USA}

\author{Giselle Galan}
\affiliation{Harvard-Smithsonian Center for Astrophysics, 60 Garden St., Cambridge, MA 02138, USA}
\affiliation{Department of Physics, Massachusetts Institute of Technology, 77 Massachusetts Avenue, Cambridge, MA 02139, USA}

\author[0000-0003-2872-2614]{Sijie Yu (\cntext{余思捷})}
\affiliation{Center for Solar-Terrestrial Research, New Jersey Institute of Technology, 323 M L King Jr. Blvd., Newark, NJ 07102-1982, USA}

\author[0000-0001-8794-3420]{Wei Liu}
\affiliation{Lockheed Martin Solar and Astrophysics Laboratory, Building 252, 3251 Hanover Street, Palo Alto, CA 94304, USA}
\affiliation{Bay Area Environmental Research Institute, NASA Research Park, Mailstop 18-4, Moffett Field, CA 94035-0001, USA}
\affiliation{W.W. Hansen Experimental Physics Laboratory, Stanford University, Stanford, CA 94305, USA}

\author[0000-0003-4695-8866]{Gang Li}
\affiliation{Department of Space Science and CSPAR, University of Alabama in Huntsville, Huntsville, AL 35899, USA}

\received{2020 July 27}
\revised{2020 Oct 9}
\submitjournal{The Astrophysical Journal}

\begin{abstract}
In this study, we investigate motions in the hot plasma above the flare loops during the 2017 September 10 X8.2 flare event.  We examine the region to the south of the main flare arcade, where there is data from the {\it Interface Region Imaging Spectrograph} ({\it IRIS}), and the Extreme ultraviolet Imaging Spectrometer (EIS) on {\it Hinode}.  We find that there are initial blue shifts of 20--60 km s$^{-1}$ observed in this region in the \ion{Fe}{21} line in {\it IRIS} and the \ion{Fe}{24} line in EIS, and that the locations of these blue shifts move southward along the arcade over the course of about 10 min.  The cadence of {\it IRIS} allows us to follow the evolution of these flows, and we find that at each location where there is an initial blue shift in the \ion{Fe}{21} line, there are damped oscillations in the Doppler velocity with periods of $\sim$400 s.  We conclude that these periods are independent of loop length, ruling out magnetoacoustic standing modes as a possible mechanism.  Microwave observations from the Expanded Owens Valley Solar Array (EOVSA) indicate that there are non-thermal emissions in the region where the Doppler shifts are observed, indicating that accelerated particles are present.  We suggest that the flows and oscillations are due to motions of the magnetic field that are caused by reconnection outflows disturbing the loop-top region.
 \end{abstract}

\section{Introduction}

The general consensus is that solar flares are powered by magnetic reconnection.  However, the details of the dynamics involved in this reconnection process are not fully understood. As the instrumentation observing solar flares improves in spatial resolution and temporal cadence, a wealth of measurements regarding the dynamics during flares has become available.

In the region where the flare current sheet is thought to be, there have been many observations of inflows \citep{Yokoyama2001,Narukage2006,Hara2011,Savage2012b,Chen2020b} and outflows \citep{Wang2007,Savage2010,Hara2011,LiuW2013,LiuR2013, Savage2012b,Takasao2012,Tian2014,Longcope2018,Hayes2019,Yu2020}, which are thought to be a direct consequence of the reconnection process. Above the flare loop tops, strong flows of hot plasma in the range of hundreds of kilometers per second have been observed in the direction perpendicular to the presumed location of the flare current sheet \citep{Innes_b2003,Imada2013,Polito2018}.  These flows have been interpreted as deflection flows in the downstream region of a termination shock, and are often accompanied by observations of broad spectroscopic lines \citep{Innes_b2003,Imada2013, Doschek2014}, a possible indication of turbulence.   

In addition to steady flows, a variety of turbulent and oscillatory behavior has also been associated with flares.  Quasi-periodic pulsations (QPPs) have been observed with a range of periods from a few seconds to a few minutes \citep[see][for a review]{VanDoorsselaere2016}.  Spectroscopic measurements have revealed oscillations in velocity and density that have been interpreted as slow magnetoacoustic waves propagating through flare loops \citep[e.g.][]{Kliem2002, WangTJ2003,Wang2011,Kumar2015,Conde2020}.  Supra-arcade fan regions have exhibited large scale turbulence \citep{McKenzie2013,Freed2018}, transverse kink waves \citep{Verwichte2005, LiL2016}, and vortex shedding behind newly reconnected loops \citep{Samanta2019}. High-cadence instrumentation such as the Atmospheric Imaging Assembly (AIA) on the {\it Solar Dynamics Observatory} ({\it SDO}) have enabled the observation of quasi-periodic fast mode magnetosonic waves propagating away from flare sites \citep{LiuW2011,ShenY2012,ShenY2018}.  High-resolution spectroscopic observations from {\it IRIS} have found evidence for global sausage oscillations \citep{Tian2016} and kink modes \citep{LiD2017b,LiD2018} in flare loops as well as oscillations in the velocities at flare ribbons \citep{Brannon2015,Brosius2015}.

Many models have been developed in order to understand the dynamics of solar flares, which show the variety of features that can influence the dynamics during a flare.  For example, termination shocks, formed where reconnection outflows impinge upon newly reconnected loops, can influence the direction and magnitude of reconnection ouflows in the region above the flare loops \citep{Forbes1986,ForbesMalherbe1991,YokoyamaShibata1998,YokoyamaShibata2001,Shiota2003,Seaton2009,Workman2011, Guo2017}.  Similarly, complicated dynamics can occur when plasmoids are formed in the current sheet due to the tearing mode instability \citep{Shibata2001,Shen2011, Yang2015,Lynch2016}.  Recently, models have begun to address the dynamics that occur when plasmoids or other reconnection outflows interact with termination shocks \citep{Nishizuka2013,Takasao2015,Takasao2016,Shen2018}.

\begin{figure*}[h]
\includegraphics[scale=0.6]{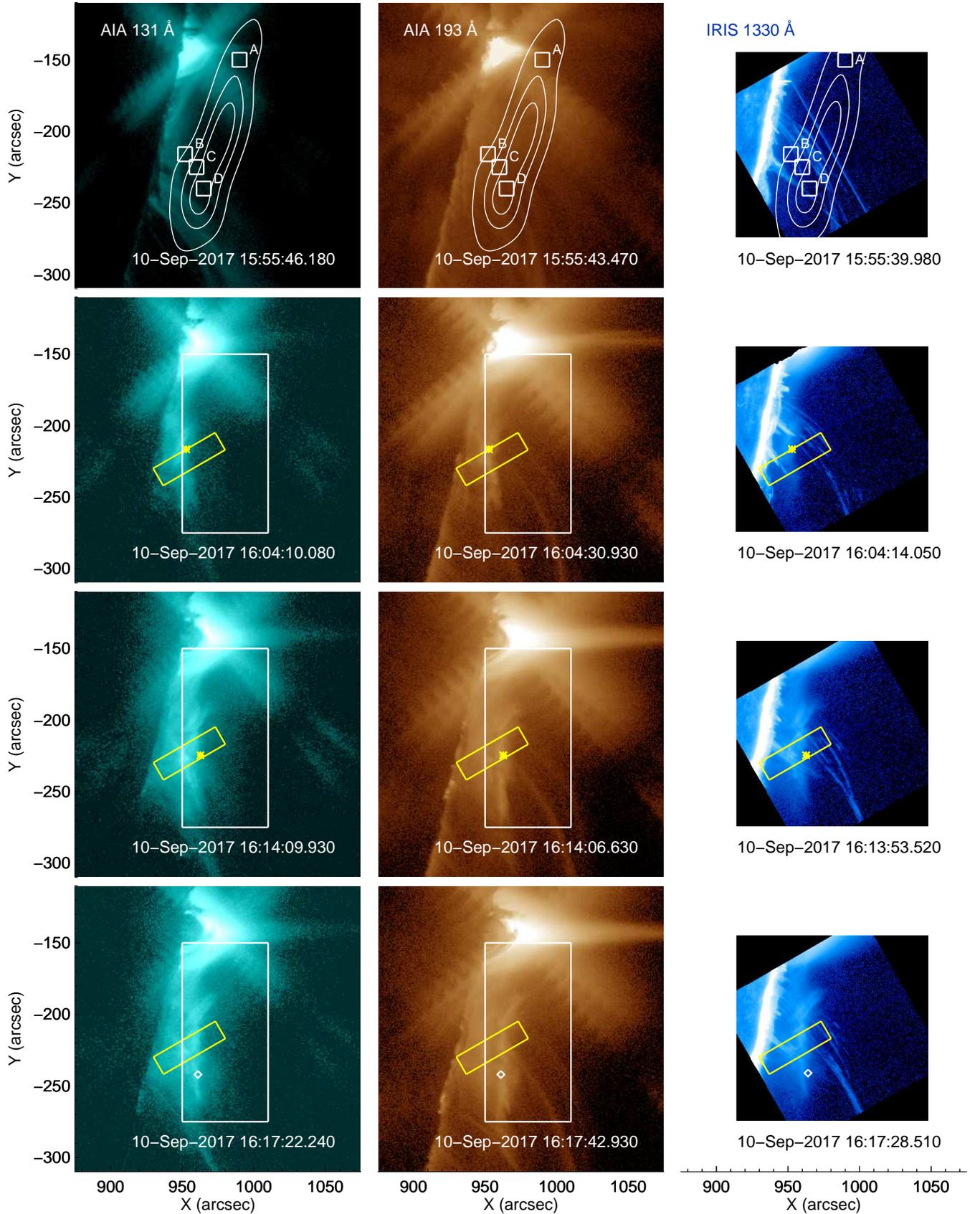}
\caption{\label{summary.fig} Images from the AIA 131 \AA\ (left), AIA 193 \AA\ (center), and {\it IRIS} SJI 1330 \AA\ (right) channels during the flare.  White contours show the EOVSA 2.9 GHz emission at 50\%, 75\%, and 90\% intensity.  The colored boxes marked A--D are the locations of the EOVSA lightcurves plotted in Figure \ref{eovsa_iris.fig}. The yellow box shows the region of {\it IRIS} slit coverage shown in Figure \ref{int_vel_nt.fig}, and the white box shows the area of the EIS slit coverage.  The yellow star shows the location of blue-shifted regions observed by {\it IRIS}, and the diamond shows the location of the blue-shifted region observed by {\it Hinode}/EIS.
}
\end{figure*}

In this work, we examine the well-studied X8.2 eruptive flare event of 10 September 2017 (SOL2017-09-10).  Many papers have been written on this event, including those studying QPPs observed in the flare \citep{Hayes2019}, dynamics in the plasma sheet region \citep{Cheng2018,Longcope2018,French2019, Yu2020}, microwave emissions from the erupting flux rope \citep{Chen2020a}, flare arcade/loop-top \citep{Gary2018,Fleishman2020,Yu2020}, and plasma sheet \citep{Chen2020b}, (E)UV spectroscopic properties of the plasma sheet and supra-arcade fan \citep{Warren2018,LiY2018,Polito2018b,Cai2019}, the eruption of the flux rope \citep{Seaton2018,Long2018,Veronig2018,Yan2018} and the global EUV wave accompanying the eruption \citep{LiuW2018,Hu2019}.  Unique to our study is an in-depth investigation of the dynamics in the flare loop-top region of the arcade to the south of the main cusp-shaped flare loops. Section \ref{observations.sec} describes the observations that we analyze.  The analysis and discussion of the results is presented in Section \ref{discussion.sec}, and the conclusions are given in Section \ref{conclusions.sec}.

\section{Observations}\label{observations.sec}

The well-studied 2017 September 10 flare occurred on the west limb, starting at about 15:50 UT.  Coordinated {\it IRIS} and {\it Hinode} observations were taken as part of {\it IRIS/Hinode} Observing Program (IHOP) 244, titled ``Joint {\it IRIS/Hinode} observations of post-eruption supra-arcade plasma.'' {\it IRIS} \citep{DePontieu2014}  observed this eruption south of the main flaring arcade with a roll of 45 degrees.  Slit jaw images were taken in the 1330 \AA\ channel with a cadence of 9.3 s. Spectra were taken in an 8 step raster, from left to right, with an exposure time of 8 s and a cadence of about 75 s per raster.  The spectral resolution of {\it IRIS} in this observation is $\sim$2.6 m\AA\ per pixel in the far ultraviolet (FUV) bandpass.  The level 2 data has been dark corrected, flat fielded, and geometrically corrected.  Most of the spectral data is off-limb, but the spectra at the bottom of the slit at the first slit position contain contributions from the limb, allowing us to use the Cl I line from these spectra to remove the orbital contribution to the wavelength calibration.

The EIS spectrometer on {\it Hinode} scanned the region using a field of view of 240\arcsec\ $\times$ 304\arcsec\ and a 2\arcsec\ wide slit with 3\arcsec\ steps between exposures. The exposure time is 5 s for each exposure and the raster took about 535 s to complete.  The spectral resolution of EIS is 22m\AA, and the spatial pixels are about 1\arcsec. EIS scanned the region from right to left over the time between 16:08 UT and 16:18 UT.
In this work, we focus on the high temperature lines from \ion{Fe}{23} and \ion{Fe}{24}, which are formed around 10--20~MK. 

The EIS level 0 data  are calibrated using the IDL routine {\tt eis\_prep},available in the SolarSoft Ware \citep[SSW;][]{Freeland1998} IDL package. There are several technical issues that need to be considered when analysing EIS data\footnote{http://solarb.mssl.ucl.ac.uk/eiswiki/Wiki.jsp?page=EISAnalysisGuide}, such as the slit tilt, which causes the line centroids to vary  with the slit position, the variation of the wavelength scale as a function of spacecraft orbit, and the uncertainty in the absolute wavelength calibration.  We corrected for the slit tilt by using the routine  {\tt eis\_slit\_tilt\_array}. The orbital variation is more difficult to remove due to the lack of the strong \ion{Fe}{12} line at 195\AA\ in this observation, and the low signal in the \ion{Fe}{12} line at 192\AA\ off limb.  
This observation has a short exposure time (5s) and the region of the flare that we are interested in extends for about 5 EIS raster steps, resulting in a total observing time of about 25~s, so we do not expect the wavelength scale drift due to the orbital period ($\sim$ 98 mins) to be significant during such a short time. There is a lack of cool reference lines in the EIS spectra that can be used to perform an accurate absolute wavelength calibration, and no quiet regions are located in the field of view, so we estimate the reference wavelength for the \ion{Fe}{24} 255\AA~line using spectra observed in the loops of the main flaring region located at around Solar-Y $\approx$-150\arcsec. This location is reasonable because the high-temperature lines in the flare loops are usually observed to be at-rest during the gradual phase of flares \citep[e.g.][]{Polito2015}. We also compare the wavelength shift in other EIS lines (such as Fe XV and Fe XVI) observed in the same EIS CCD as the Fe XXIV 255\AA~line. This method yields some uncertainty ($\approx$~5--8 km s$^{-1}$), which we estimate based on the scatter of values we obtain from different lines. These values are reasonable considering the uncertainties typically associated with the EIS wavelength calibration, which are at best $\approx$~4--5 km s$^{-1}$ \citep[e.g.][]{Young2012}.

\begin{figure*}
\includegraphics[scale=0.35]{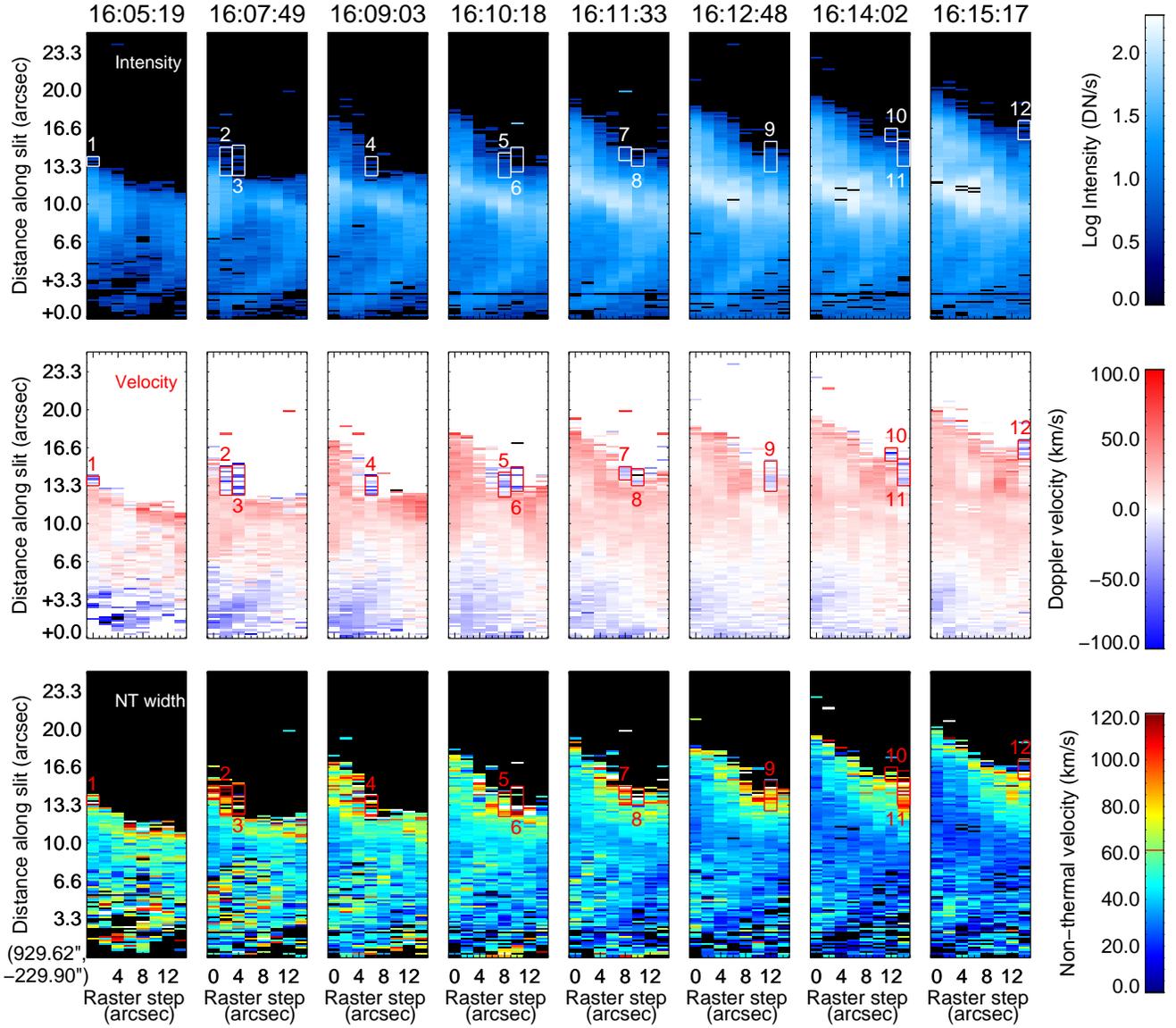}
\caption{\label{int_vel_nt.fig}  The intensity (top row), the Doppler velocity (middle row) and the non-thermal velocity (bottom row) for pixels along the slit in 8-step rasters, corresponding to the yellow box shown in Figure \ref{summary.fig}.  The time given in the top row is the time for the first raster step.  Color bars for each row are shown to the right.  The field of view is the yellow box shown in Figure \ref{summary.fig}.  Boxes show the pixels that are averaged together to get average Doppler shifts in Figure \ref{oscillations.fig}. 
}
\end{figure*}

The X-Ray Telescope \citep[XRT,][]{Golub2007} observed the eruption with high cadence ($\sim$6 s), full resolution Be-med filter images, interspersed with occasional Al-poly (binned 2$\times$2), Al-med/Al-thick and Be-thick images.  Full-resolution XRT images have a resolution of 1.0286 arcsec per pixel.  XRT observed the early part of the eruption, but does not have images between 16:04:20 UT and 16:18:00 because the flare-finding algorithm caused readout of a region of the CCD that was too far south. For this study, we use Be-med images between 16:03:50 UT and 16:04:20 UT.  Images have been calibrated using {\tt xrt\_prep}, which removes vignetting, does the dark calibration, and removes the CCD bias \citep{Kobelski2014}.  We use calibrated effective areas from the calibration presented in \citet{Narukage2014} to generate temperature response functions.

For this observation, AIA images are taken every 12 s in the six Fe dominated EUV channels with alternating long and short exposure times, and images in 1600 \AA\ and 1700 \AA\ are taken every 24 s.  The AIA data are calibrated using the {\tt aia\_prep} routine, available in SSW.  This routine aligns the channels from the four AIA telescopes, accounts for rotation in the images, and re-factors the images so that they all have the same plate scale of 0.6\arcsec per pixel.  

The Expanded Owens Valley Solar Array (EOVSA) took data of the full solar disk (including this region of interest) from $\sim$14:30 UT to 01:10 UT of the next day. An overview of the EOVSA observations of this event in different flare phases and initial imaging spectroscopy results are discussed in \citet{Gary2018}. More detailed discussions on calibration and imaging strategies are available in \citet{Chen2020a}. Briefly, EOVSA obtained data of this event in 2.5--18 GHz with 134 frequency channels spread over 31 equally spaced spectral windows (SPWs). Each SPW has a bandwidth of 160 MHz. The center frequencies of these SPWs are given by $\nu=2.92 + n/2$ GHz, where $n$ is the SPW number from 0 to 30. Images were made by combining the spectral channels within each of the 31 spectral windows using the CLEAN algorithm. A circular beam with a size of $73''.0/\nu_{\rm GHz}$ is used for restoring the CLEAN images below 14.5 GHz. The nominal full-width-half-max (FWHM) angular resolution is $113''.7/\nu_{\rm GHz}\times53''.0/\nu_{\rm GHz}$ at the time of the observation. 

A summary of the observations is shown in Figure \ref{summary.fig}.  The main eruption occurred at about 15:50 UT.  We will focus on the area to the south of the main eruption and bright cusp-shaped flare loops that are visible at the top of the AIA images in Figure \ref{summary.fig}.  Shortly after the eruption, at 15:55 UT, the EOVSA 2.9 GHz emission extends south from the bright flare loops, as seen in the top panel of Figure \ref{summary.fig}.  At this time, there is some faint emission in the AIA 131 \AA\ channel close to the limb, and some strands of emission in the {\it IRIS} 1330 \AA\ channel that look similar to coronal rain.  As the eruption progresses (lower panels of Figure \ref{summary.fig}), faint, diffuse emission is seen extending to the south of the bright flare arcade in the AIA 131 \AA\ and 193 \AA\ channels (and also in the AIA 94 \AA\ and 335 \AA\ channels, which are not shown), and in the {\it IRIS} 1330 \AA\ SJI image.

\subsection{Spectral parameters from {\it IRIS} and EIS \label{spectral.sec}}

For this observation, the {\it IRIS} slit is in the perfect position to document the dynamics of the emission extending to the south of the main bright flare arcade.  {\it IRIS} was pointed off-limb, and the spectra show that the emission seen in {\it IRIS} early in the eruption is largely from the \ion{Fe}{21} line, which is sensitive to $\sim$10 MK plasma, though there is also some cool coronal rain visible that appears as emission in the \ion{C}{2} lines.  We fit a Gaussian to the \ion{Fe}{21} line in the spectra at every raster position in the yellow box shown in Figure \ref{summary.fig}.  The intensities, Doppler velocities, and non-thermal widths in this region are shown in Figure \ref{int_vel_nt.fig} for eight consecutive rasters covering times between 16:05 UT and 16:16 UT.  

The plot of the Doppler shifts (middle row in Figure \ref{int_vel_nt.fig}) shows that the diffuse emission seen in the {\it IRIS} SJI images is largely blue shifted at the bottom of the yellow box shown in Figure \ref{summary.fig}, and red shifted towards the top.  In some locations, there are prominent blue shifts that appear above the red shifted emission.  The locations of these blue shifts (indicated by yellow stars in Figure \ref{summary.fig}) are at or near the top of the diffuse emission seen in the AIA 131 \AA\ and the {\it IRIS} SJI images.  We speculate that this blue shifted emission is at or above the tops of newly formed, hot loops, and we will therefore refer to these features as ``loop-top blue shifts.''  We will substantiate this speculation below in section \ref{cooling_loops.sec}.

\begin{figure*}
\includegraphics[scale=0.55]{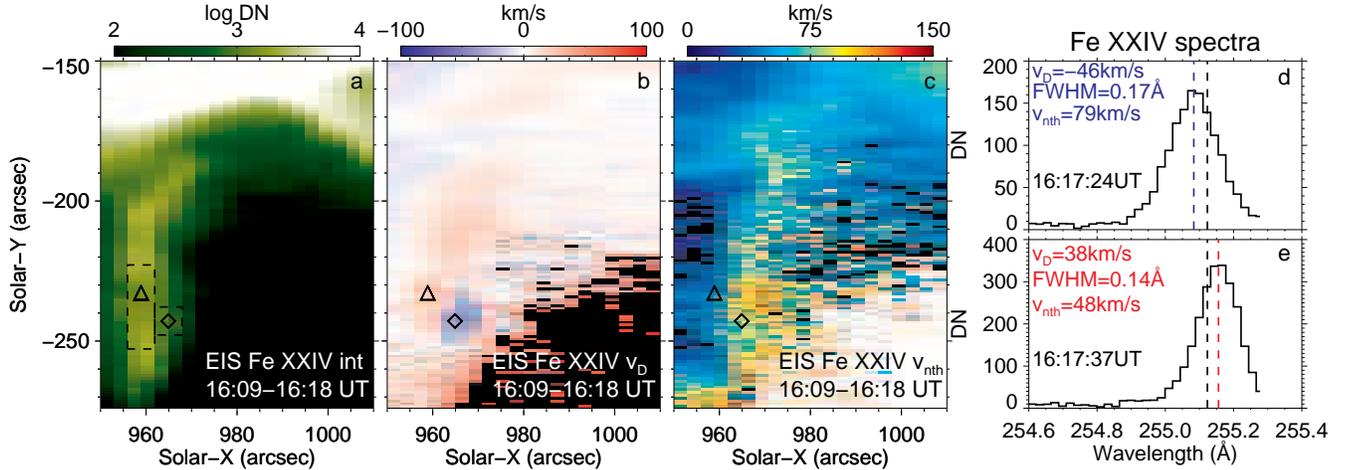}
\caption{\label{eis_map.fig} Panel a) shows the intensity of the EIS \ion{Fe} {24} line, panel b) shows the Doppler velocity, and panel c) shows the non-thermal velocity for spectra in the EIS raster that was taken between 16:09 and 16:18 UT.  Panels d) and e) show \ion{Fe}{24} line profiles from the locations marked by the diamond and triangle, respectively.}
\end{figure*}

The non-thermal widths of the \ion{Fe}{21} line are shown in the bottom row of Figure \ref{int_vel_nt.fig}.  We note that the pixels at the top have broader non-thermal widths than the rest of the emission, especially for the later rasters.    This result is consistent with previous observations that have found that non-thermal widths tend to increase with height in the supra-arcade region \citep{Doschek2014}. We note that an increasing non-thermal width with height was also reported by both \citet{Warren2018} and \citet{LiY2018} for the plasma sheet above the central flare arcade in the same event, which has a nearly edge-on viewing perspective.

The Gaussian parameters from the EIS raster of the region, shown in Figure \ref{eis_map.fig}, have similar features.  There is only one EIS map for the time period of interest, since it takes EIS almost 10 minutes to scan the region. Figure \ref{eis_map.fig}a shows the intensity in the EIS \ion{Fe}{24} line.  Similar to the AIA and {\it IRIS} SJI images, a structure is visible in the \ion{Fe}{24} intensity extending to the south of the bright flare loops, indicated by the dashed-line boxes in Figure \ref{eis_map.fig}a.  The plot of the Doppler shift of the EIS \ion{Fe}{24} line (Figure \ref{eis_map.fig}b) shows that this emission is mostly red shifted, except for one location that shows prominent blue shifts. The location with prominent blue shifts is marked by a black diamond in Figure \ref{eis_map.fig}a-c, and a white diamond shows the same position relative to the AIA and {\it IRIS} images in Figure \ref{summary.fig}. We emphasize that, despite the EIS Doppler shifts being affected by a significant uncertainty (as discussed in Section \ref{observations.sec}), the values measured here and the relative shifts between different regions of the flare arcade are well above this uncertainty, as shown in Figures \ref{eis_map.fig}d and \ref{eis_map.fig}e. Similar to the {\it IRIS} \ion{Fe}{21} results, the plot of the non-thermal widths of the EIS line (Figure \ref{eis_map.fig}c) shows an area of enhanced non-thermal velocity at the top edge of the \ion{Fe}{24} emission.

\begin{figure*}
\includegraphics[scale=0.6]{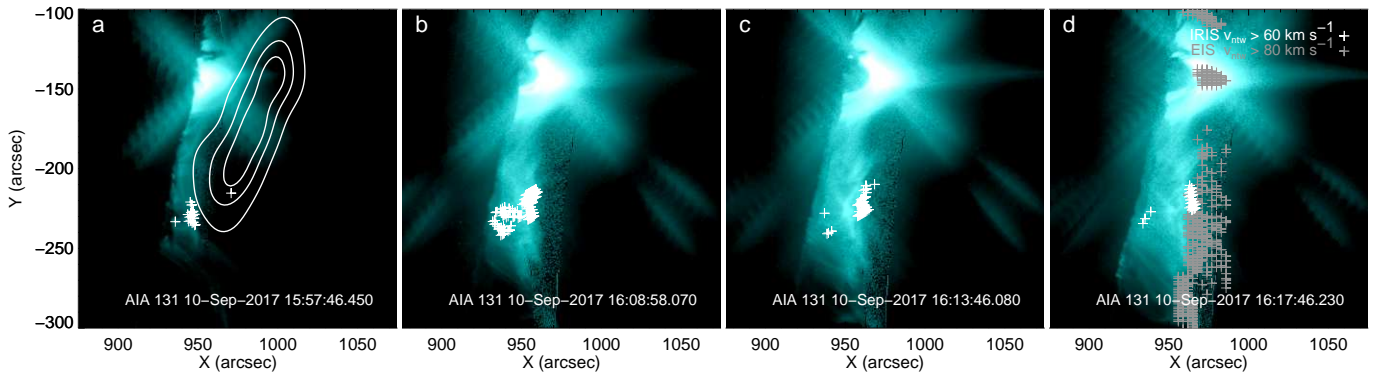}
\caption{\label{nt_vel_aia.fig} Positions of {\it IRIS} \ion{Fe}{21} nonthermal widths $>$ 60 km s$^{-1}$ (white + signs) and EIS \ion{Fe}{24} nonthermal widths $>$ 80 km s$^{-1}$ (grey + signs) over the time span of an {\it IRIS} raster (75 s) plotted on top of AIA 131 \AA\ images at four different times.  Contours in panel a) show EOVSA 2.9 GHz emission at 50\%, 75\%, and 90\% intensity. (An animation of this figure is available).}
\end{figure*}

Figure \ref{nt_vel_aia.fig} shows the locations of large non-thermal velocities from EIS and {\it IRIS} plotted on top of AIA 131 \AA\ images. Early in the eruption at 15:57:46 UT (Figure \ref{nt_vel_aia.fig}a), large non-thermal widths in the {\it IRIS} \ion{Fe}{21} line are seen at the top of the diffuse AIA 131 \AA\ emission that extends southward from the main flare arcade.  As the eruption progresses, large non-thermal widths in this line spread throughout the diffuse structure seen in the AIA 131 \AA\ images (Figure \ref{nt_vel_aia.fig}b). After about 16:13 UT, the persistently high non-thermal velocities in the {\it IRIS} \ion{Fe}{21} line remain just above the brightest emission in that part of the 131 \AA\ images, but are not as prevalent closer to the limb (Figure \ref{nt_vel_aia.fig}c).  When the EIS slit reaches this region at about 16:17 UT (Figure \ref{nt_vel_aia.fig}d), the \ion{Fe}{24} line also exhibits high non-thermal widths just above the brightest 131 \AA\ emission in the southern part of the image.

\begin{figure*}
\includegraphics[scale=0.6]{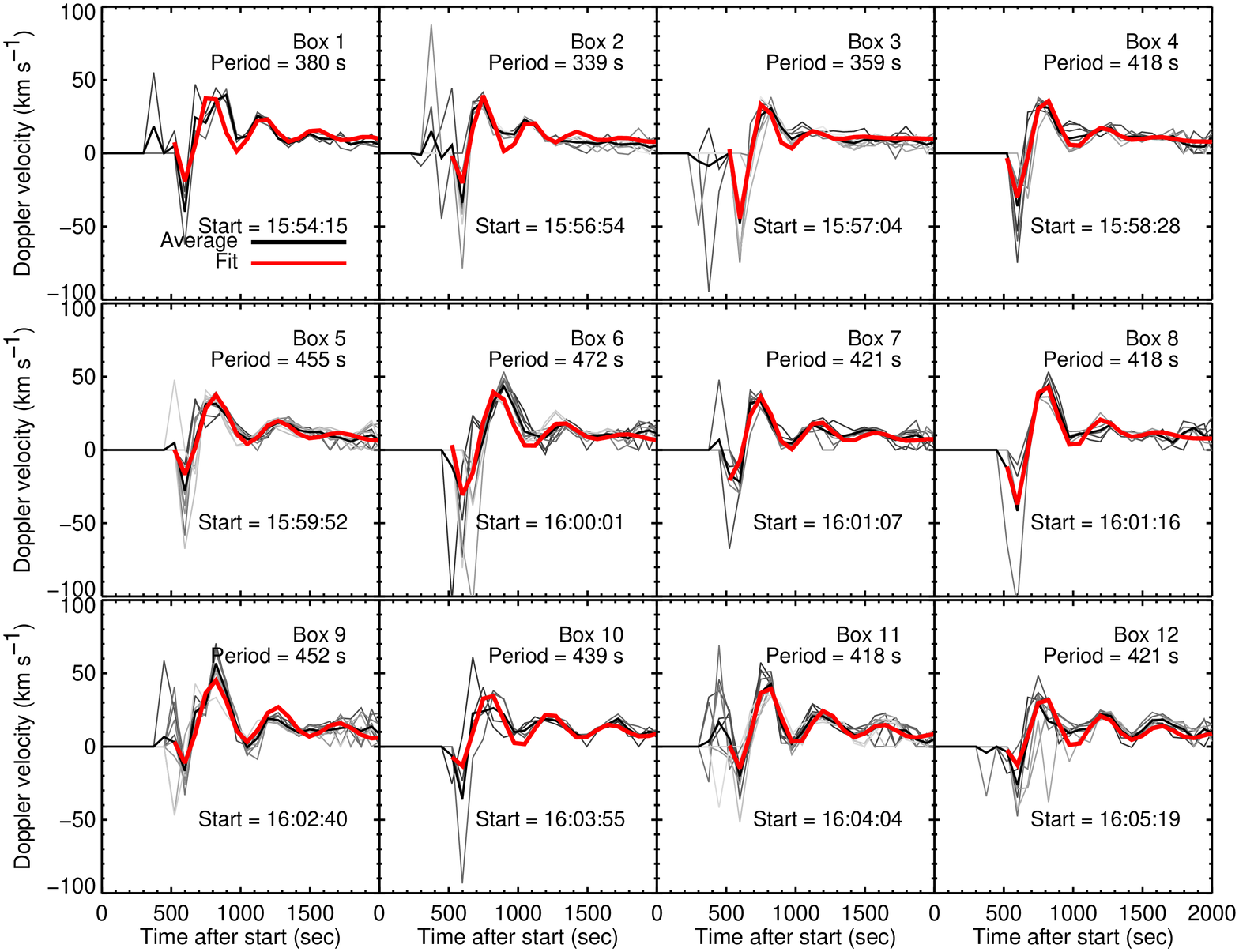}
\caption{\label{oscillations.fig}  {\it IRIS} Doppler shifts as a function of time for the pixels in the boxes shown in Figure \ref{int_vel_nt.fig}.  Times have been shifted so that the main blue shift for each time series occurs at the same time. For each plot, thin gray lines indicate the time series for each pixel in the box, averages for all the pixels in the box are shown as a thick black line, and a fit to the average using a damped oscillation is shown as a thick red line.}
\end{figure*}

Because the {\it IRIS} rasters take less time than the EIS rasters ($\sim$75 s each) there are repeated scans of the region of interest, allowing us to examine the behavior of the loop-top Doppler shifts as a function of time.  Figure \ref{oscillations.fig} shows time series of the Doppler shift for the pixels in each box in Figure \ref{int_vel_nt.fig}.  We average the Doppler shifts observed in all of these pixels, and fit the resulting curve to a damped oscillator.  We find that the periods of these oscillations are all fairly similar, and are between 340 and 470 s.  We note that the cadence of the {\it IRIS} observations at each location is about 75 s, and thus the oscillations in the Doppler shift could have a shorter period that is not detected due to the sampling frequency at each location.  For example, all of the oscillations shown in Figure \ref{oscillations.fig} can also be fit with oscillations with shorter periods of 61-64 s.  Interestingly, \citet{Hayes2019} find QPPs in this same event with a period of about 65 s between 15:50 to 16:15 UT, which overlaps with our time period of interest, though the location of these QPPs is interpreted to be in the main flare arcade.  Higher cadence spectroscopic observations would be needed to determine if there is indeed a shorter period in the Doppler shift observations.

\subsection{Ribbon motion observed in AIA}

\begin{figure*}
\includegraphics[scale=0.6]{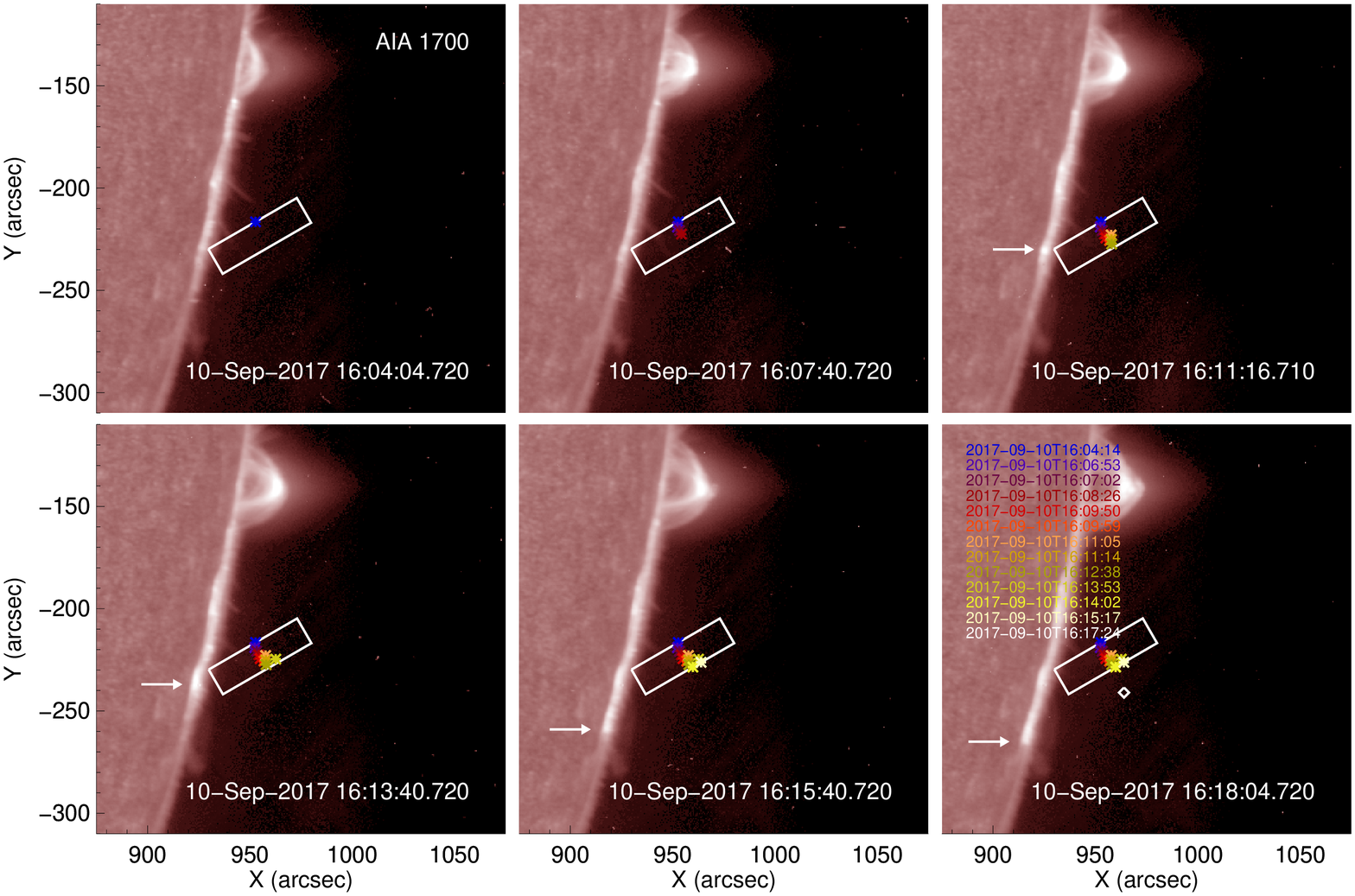}
\caption{\label{uv_summary.fig} AIA 1700 \AA\ images showing footpoint spreading to the south indicated by arrows. The {\it IRIS} raster field of view and the locations of the blue-shifted loop-top spectra (i.e. the midpoints of the boxes shown in Figure \ref{int_vel_nt.fig}) are also shown.  (An animation of this figure is available).
}
\end{figure*}

This flare occurred right on the limb, and one of the ribbons of the flare arcade is visible as the flare progresses.  Figure \ref{uv_summary.fig} shows images from the AIA 1700 \AA\ channel, with the locations of the loop-top blue shifts plotted as colored asterisks.  From these images and the movie accompanying Figure \ref{uv_summary.fig}, it is clear that the flare ribbon is spreading to the south at the same time that the loop-top blue shifts are seen moving south in the {\it IRIS} field of view.  Previous studies of the active region while it was on the disk have found that there was a significant neutral line extending approximately from north to south between two strong patches of magnetic flux \citep[e.g.][]{Hou2018} along with a pre-existing reverse S-shaped filament \citep{Chen2020a}.  This configuration suggests that the intensity increase southward seen in the AIA 1700 \AA\ images is an intensity front moving along the ribbon.

\subsection{Loop-top microwave source observed by EOVSA}

\begin{figure*}
\includegraphics[scale=0.6]{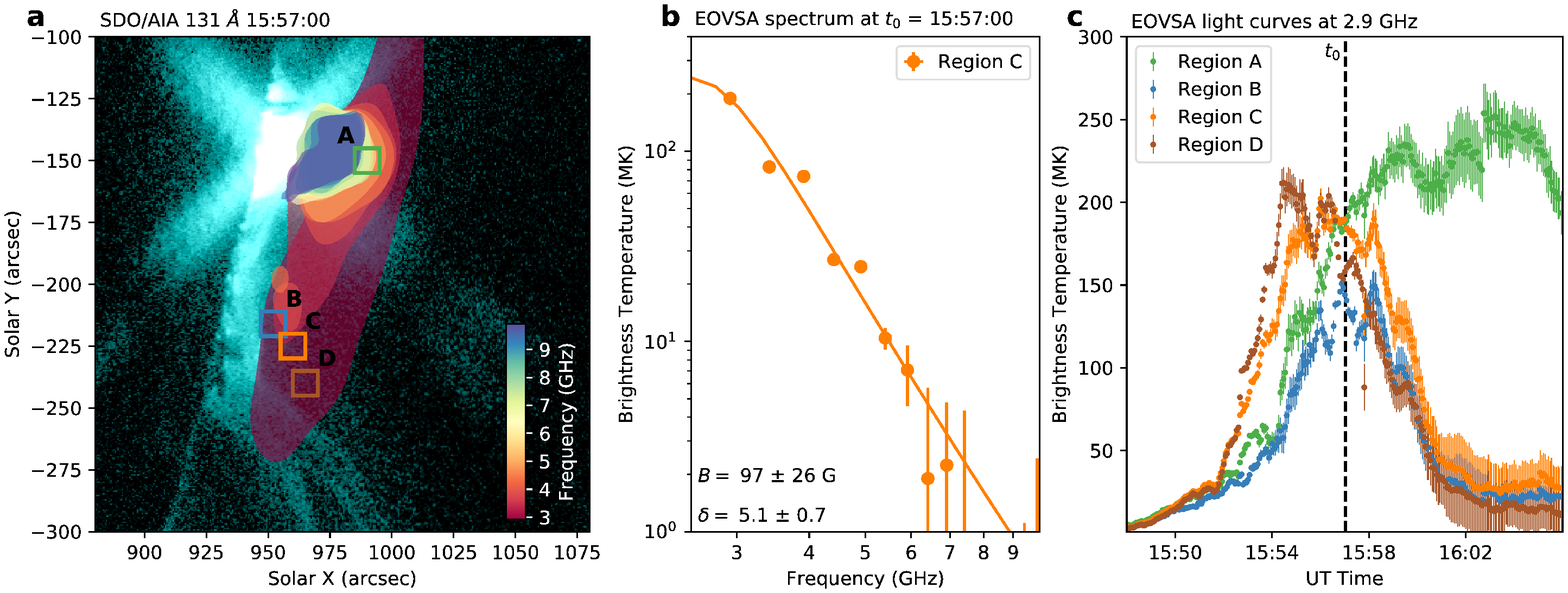}
\caption{\label{eovsa_iris.fig} Panel a: Contours of EOVSA microwave emission as a function of frequency observed at 15:57 UT (40\% of the maximum at the respective frequency) superimposed on an AIA 131 \AA\ image. Boxes B-D indicate locations where blue shifts are observed in the {\it IRIS} \ion{Fe}{21} line a few minutes later.  Box A indicates the location of the flare arcade, for reference. Panel b: The spectrum of the microwave emission from box C (symbols) and a fit (solid line). Panel c: Microwave brightness temperature as a function of time for the emission from boxes A--D. Vertical dashed line indicates the timing of the microwave image and spectrum shown in panels a and b.}
\end{figure*}

Figure \ref{eovsa_iris.fig}a shows contours of microwave emission as a function of frequency at 15:57 UT, plotted over an AIA 131 \AA\ image from the same time.  Boxes on the image indicate locations of the main flare arcade (box A) and the locations of the observed loop-top blue shifts (boxes B--D). In Figure \ref{eovsa_iris.fig}c, we show the microwave brightness temperature light curves for several locations indicated on the {\it IRIS} image.  The peak brightness temperature at 2.9 GHz is very large ($>$100 MK), indicating the presence of nonthermal emission at these locations \citep{Chen2020a}. An example of a spatially resolved microwave spectrum at 15:57 UT is shown in panel b of Figure \ref{eovsa_iris.fig} for box C. The spectrum displays a power-law shape with a negative spectral index, which is characteristic of optically-thin gyrosynchrotron radiation. Spectral analysis (using the techniques described in \citealt{Fleishman2020} and references therein) suggests that the magnetic field strength in this region is on the order of $\sim$100 G. The relatively weak magnetic field (compared to those in the central flaring region in, e.g., \citealt{Fleishman2020,Chen2020b}) combined with the soft electron distribution (spectral index $\delta \approx 5.1\pm 0.7$) are likely responsible for the microwave spectrum peaking at a low frequency (at or below 2.9 GHz). However, EOVSA (or RHESSI) images do not show a loop-top source at the same time and location as the loop-top blue shifts observed with {\it IRIS} and EIS. The absence of the loop-top source in the microwave image at this time is possibly due to the instrumental dynamic range, which would be a limiting factor in detecting an already diminished source after $\sim$16:01 UT given the presence of the very bright source in the main cusp-shaped loop to the north (see Figure \ref{eovsa_iris.fig}c).

\section{Analysis and Discussion} \label{discussion.sec}
\subsection{Flare Loops and Cooling Times \label{cooling_loops.sec}}

In Section 2, we speculate that the initial strong blue shifts located in the diffuse emission in the {\it IRIS} SJI images are located at or near the tops of newly formed loops. In order to determine if this assumption is a reasonable one, we assume that during the early phase of the flare the cooling is primarily due to conduction, but radiative cooling dominates later in the event, when the flare loops appear in the {\it IRIS} SJI images.  Thus we estimate the cooling times of possible loops using the formula \citep[e.g.][]{Cargill1995}
\begin{equation}
\label{cargill.eq}
\tau_{cool} = 0.0235\frac{L^{5/6}}{(Tn)^{1/6}}
\end{equation}
where $T$ and $n$ are the initial temperature (in K) and density (in cm$^{-3}$), respectively, in the loop, and $L$ is the loop half-length (in cm).   To get $L$, we assume that the loops are semi-circular, and that the measured height above the limb is the loop radius.  We will use two methods to get $T$ and $n$, as described below.  Equation \ref{cargill.eq} holds for evaporative (rather than static) conductive cooling, and assumes that the initial conductive cooling time is much shorter than the initial radiative cooling time.  This assumption is reasonable given that the first indication of these loops is in the AIA channels that observe emission hotter than 10 MK, a temperature at which conductive cooling times are short.

\begin{figure*}
\includegraphics[scale=0.48]{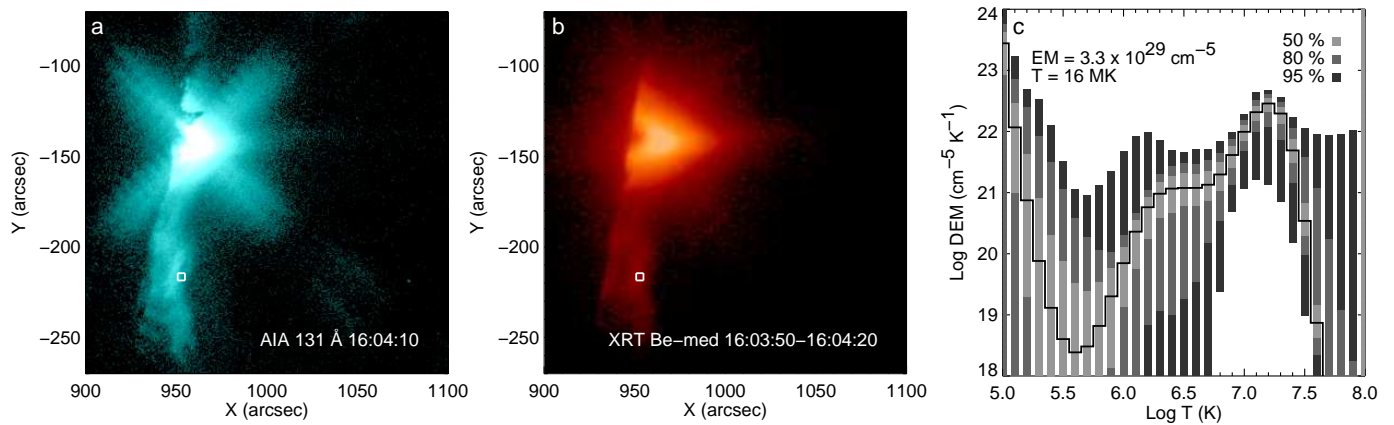}
\caption{\label{saf_dem.fig}  Panel a): AIA 131 \AA\ image used in the DEM calculation.  Panel b): XRT Be-med image made from averaging six images between 16:03:50 and 16:04:20. Boxes on the left and middle panels indicate the location of the pixels averaged to calculate the DEM.  Panel c): DEM (solid line) calculated at the indicated location on the images.Dark gray, gray and light gray boxes on the DEM plot encompass 95\%, 80\% and 50\% of the Monte Carlo solutions, respectively.  }
\end{figure*}

In order to estimate the plasma temperature and density, we perform a differential emission measure (DEM) calculation with the routine {\tt xrt\_dem\_iterative2} \citep{Weber2004,Golub2004,Cheng2012} using data from the six Fe-dominated AIA EUV channels and the XRT Be-med filter. This routine has been used previously to determine temperatures and emission measures in the supra-arcade fan region of flares \citep{Hanneman2014,Reeves2017}.  The results of the DEM calculation are shown in Figure \ref{saf_dem.fig}.  The signal in the XRT images is low, so we average six images taken between 16:03:50 UT and 16:04:20 UT.  We spatially average the signal in the six AIA filters and the time-averaged XRT image in a box centered on the location of the first loop-top blue shift observed with {\it IRIS}.  The error in the observed intensities is taken as the standard deviation of intensities in the box in each filter.  We calculate 1000 Monte Carlo iterations of the DEM using the measured intensities varied by a normally distributed random error.  The right panel of Figure \ref{saf_dem.fig} shows the DEM, with bars encompassing 50\%,80\% and 95\% of the solutions.  There is a clear peak at about 16 MK, and the emission measure is 3.3$\times10^{29}$ cm$^{-5}$.  In order to calculate the density, we use the equation
  \begin{equation}
  \label{dens.eq}
n_e \approx \sqrt{\frac{EM_h}{0.83 \cdot h}}
    \end{equation}
where $EM_h$ is the line of sight emission measure calculated by integrating the DEM over $T$, and we assume a fully ionized gas with helium abundance relative to hydrogen equal to 0.1 such that $n_e n_H = 0.83n_e^2$. The variable $h$ is the depth of the plasma column, which we take to be 10\arcsec.6, following \citet{Warren2018}.  Using this equation, we obtain a density of 2.3$\times10^{10}$ cm$^{-3}$.  \citet{Cai2019} performed a DEM of a similar time and location, and found much lower values for the temperature and emission measure (9.65 MK and 7.99$\times10^{27}$ cm$^{-5}$, respectively), possibly because they only used the AIA filters, and did not include the XRT, which is sensitive to hotter plasma.

In order to verify our values for $T$ and $n$, we use the ratio of the temperature sensitive \citep[see e.g.][]{Polito2018b} EIS Fe XXIII 263.76\AA~and Fe XXIV 255.10\AA~lines, at the time and location of the loop-top blue shift observed in EIS. The Fe XXIII and Fe XXIV line intensities are converted from data numbers to physical units using the radiometric calibration of \cite{DelZanna2013b}. In the same location and time, we estimate the electron number density by calculating the emission measure of the Fe XXIV plasma that dominates the signal in the AIA 193\AA~images, following the method detailed in \cite{Polito2018b}. The emission measure can be expressed as the ratio of the intensity in that filter and the filter response function at the $T$ calculated using the EIS temperature diagnostics described above. We assume coronal abundances, and a filling factor of 1. Once the emission measure is known, the density can be obtained using Equation \ref{dens.eq}.  With this method, we find that the temperature is about 14.5 MK, and the density is 2.0$\times10^{10}$ cm$^{-3}$. These values are similar to the temperatures and densities found in the main plasma sheet above the cusp-shaped loops by \citet{Warren2018} using a similar method, and also agree very well with the values we obtain via the DEM calculation. 

\begin{figure*}
\includegraphics[scale=0.48]{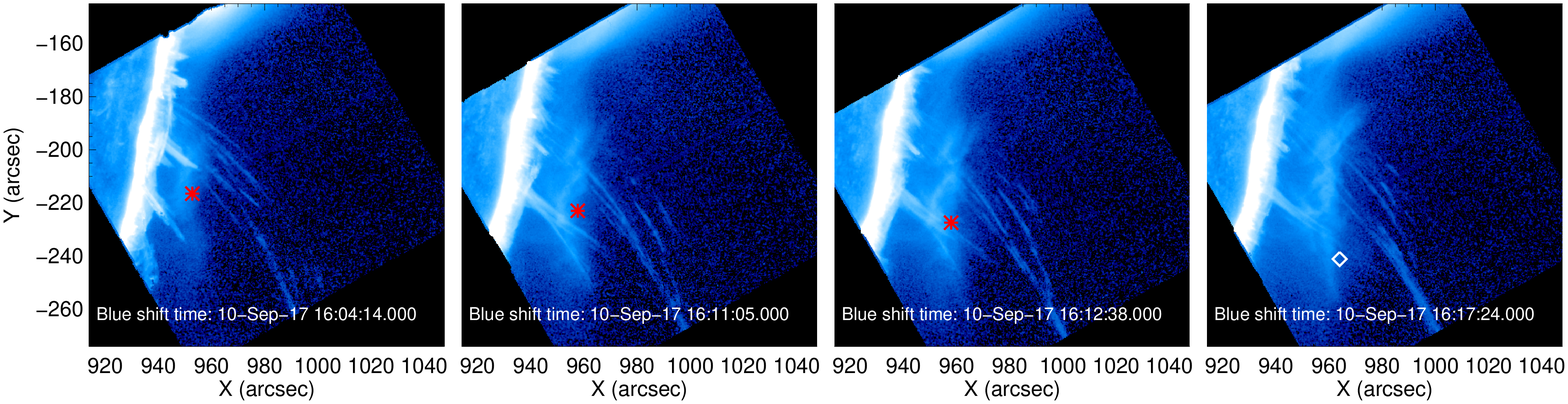}
\includegraphics[scale=0.48]{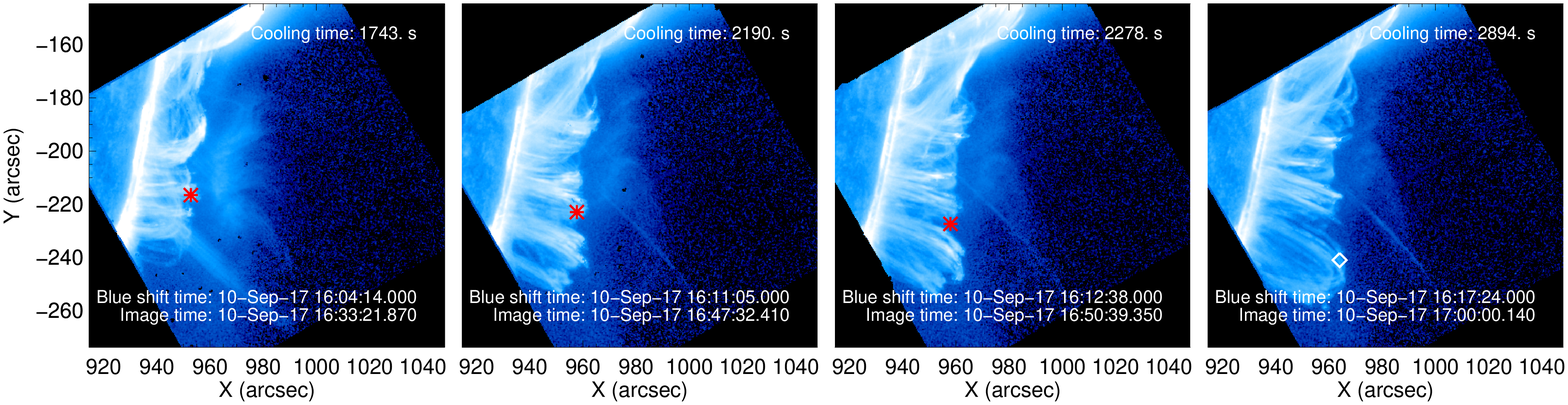}
\caption{\label{flare_loops_cooling.fig}  Top: Locations of selected blue shifts shown on an {\it IRIS} 1330 \AA\ SJI image at the time of their occurrence ($t_0$).  Bottom:  The same locations, plotted on an {\it IRIS} 1330 \AA\ SJI image at the time nearest to $t = t_0 + \tau_c$, where $\tau_c$ is the cooling time. }
\end{figure*}

Since our values of temperature and density calculated with two different methods (DEM and EIS ratio+AIA 193~\AA~EM) are similar, we use the values of temperature and density derived from the DEM as reasonable estimates for all the loops, and we use these values to calculate the cooling time, $\tau_c$, for the loops. In the top panels of Figure \ref{flare_loops_cooling.fig}, we show the locations of some of the ``loop-top'' blue shifts plotted on the {\it IRIS} SJI 1330 \AA\ image closest in time to when the blue shift occurred.  In the bottom panels, we show the same blue shift locations superimposed on an {\it IRIS} SJI 1330 \AA\ image advanced in time by $\tau_c$.  In these images, the locations of the blue shifts correspond quite nicely to the tops of the loops that appear in the {\it IRIS} 1330 \AA\ SJI images that are due to emission from cool lines such as \ion{C}{2}.  This result is strong evidence in favor of the hypothesis posed in Section \ref{spectral.sec} that the Doppler shifted emission we observe with {\it IRIS} is near the tops of newly formed loops.

\begin{figure}
\includegraphics[scale=0.4]{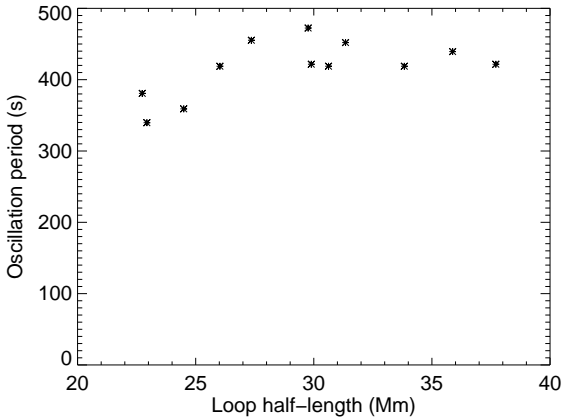}
\caption{\label{period_height.fig}  Period of the oscillation observed in the {\it IRIS} \ion{Fe}{21} line as a function of loop half-length for the midpoints of each of the boxes shown in Figure \ref{int_vel_nt.fig}.  The loop half-length is estimated by assuming that the loops are semi-circular, and that the measured height above the limb is the loop radius.}
\end{figure}

\subsection{Origins of loop-top Doppler shifts}
Flows observed at or above the tops of flare loops have previously been interpreted as deflection flows from a termination shock \citep[e.g.][]{Innes_b2003,Imada2013,Polito2018} and recent work by \citet{Cai2019} showed that the diffuse emission visible in the {\it IRIS} SJI images of this event is a probable location for a termination shock.  However, we observe damped oscillations in the Doppler shift, which are not consistent with the idea of these velocities being simple deflection flows.  \citet{WangTJ2003} find similar damped oscillations in the Doppler shifts with periods of 420-1860 s in hot flare lines ($T >$ 6 MK) from SoHO/SUMER data.  They interpret these oscillations as a standing slow mode wave caused by the injection of hot plasma at one footpoint.  Some aspects of this interpretation fit our data quite well: the periods are similar, the oscillations are damped, and there are corresponding brightenings at the footpoints seen in the AIA 1700 \AA\ data.  There is one aspect of this interpretation that does not fit our observations, however, which is the relationship of the oscillation period to the loop length.  The period of the slow standing mode is related to loop length in the following manner:
\begin{equation}
\tau_s = \frac{4L}{jc_t}
\end{equation}
 where $L$ is the half loop length as before, $c_t$ is the phase speed (which is approximately the sound speed), and $j$ is a mode number that can be taken as 1 or 2, as these modes are the easiest to excite \citep{Roberts1984,Nakariakov2004}.  Figure \ref{period_height.fig} shows the period of our oscillation as a function of loop half-length, where the loop half-length is estimated as by assuming that the loops are semi-circular, and that the measured height above the limb is the loop radius, as before.  There is no clear relationship between the loop half-length and the period of the oscillation, so the oscillations can not be due to the standing slow mode.
 
Another possible explanation for oscillations at the flare loop tops was given by \citet{Takasao2016}, who found that colliding shock fronts due to reconnection outflows in 2D MHD simulations resulted in oscillations with periods of tens to hundreds of seconds. The oscillations disappear when the shock front becomes horizontal.  Similarly, modeling done by \citet{Cai2019} found that plasmoids impacting the loop tops could cause significant transverse oscillations with a period of about 50 s.  In both of these models, the oscillations in the velocity are expected to appear near the top or above the increase in density due to the outflows impinging on the loops below.  In our observations, the Doppler flows are located near the top of the diffuse supra-arcade emission seen in the {\it IRIS} SJI and AIA 131 and 193 \AA\ images (see Figure \ref{summary.fig}.)  We note that the oscillations shown in Figure \ref{oscillations.fig} have similar amplitudes, which would require impacting plasmoids to have similar momenta in the scenario suggested by \citet{Cai2019}.

\begin{figure*}
\includegraphics[scale=0.53]{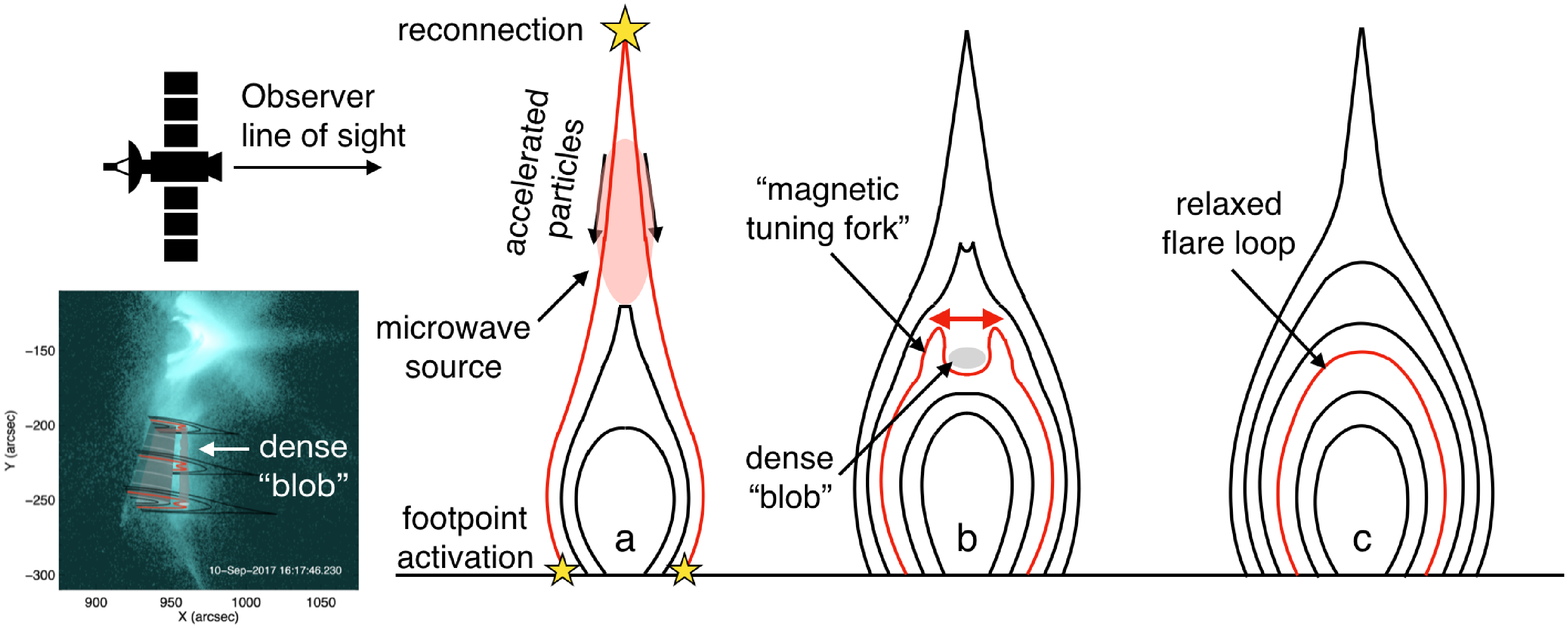}
\caption{\label{reconnection.fig}A cartoon showing a possible scenario to describe the observed oscillations. Drawing a shows the early phase of the eruption, when the microwave source is visible, as in Figure \ref{eovsa_iris.fig}.  Drawing b shows the formation of the ``magnetic tuning fork'' caused by reconnection outflows hitting the closed loops below and causing the Doppler oscillations shown in Figure \ref{oscillations.fig}.    Drawing c shows the relaxed state of the flare arcade, when the oscillations have stopped and the loops have cooled, as in the bottom row of Figure \ref{flare_loops_cooling.fig}.  The AIA 131 \AA\ image at the lower left shows the orientation of the reconnecting arcade and the location of the dense ``blob''.}
\end{figure*}

A potential scenario explaining our observations based on the modeling of \citet{Takasao2016} is shown in Figure \ref{reconnection.fig}.  First, magnetic fields are reconnected and electrons are accelerated, creating a microwave source at the base of the current sheet with a nearly face-on viewing perspective (as described in detail in \citet{Chen2020a}) and causing brightenings at the flare ribbons.  As the reconnection outflows impinge on the loops below, they create a horn-shaped magnetic field line that oscillates due to the rebounding reconnection outflow (or ``backflow''), creating magnetoacoustic waves.  \citet{Takasao2016} refer to this structure as a ``magnetic tuning fork,'' because of its similarity to a musical tuning fork that vibrates and generates sound waves.  Oscillations in this region cause the Doppler oscillations that we observe in the {\it IRIS} data.  Additionally, a dense blob of material is created within the arms of the magnetic tuning fork due to the compression of the reconnection flow.  We speculate that if the magnetic structure depicted Figure \ref{reconnection.fig} were turned to the side and elongated, this dense blob could correspond to the extended diffuse loop-top emission observed to the south of the main arcade in the AIA 131 and 193 \AA\ images and the {\it IRIS} SJI images.  This orientation and the possible location of the blob is indicated in the lower left panel of Figure \ref{reconnection.fig}.

The  \citet{Takasao2016} model is a good comparison for the observed flare because the physical parameters used in the model are comparable to the observations.  The heights of the reconnected loops are observed to be on the order of 30 Mm (see Figure \ref{period_height.fig}), and previous studies of this flare have shown that the reconnection point is on the order of 50 Mm at around 16:20 UT \citep{Yu2020}, which are similar length scales to the model.  Additionally, we calculate the reconnection rate for this event using the ratio of the inflow (18-22 km s$^{-1}$) and outflow (350-800 km s$^{-1}$) velocities previously measured for this flare at 16:00 -- 16:10 UT \citep[Figures 2 and 3 in][]{Cheng2018}. This calculation gives an estimate of 0.02--0.06 for the reconnection rate, which encompasses  the modeled value of 0.06.

According to \citet{Takasao2016}, the period of the above the loop oscillations increases as a function of the plasma $\beta$, and the maximum velocity of these motions decreases as a function of $\beta$.  In order to determine $\beta$, it is necessary to determine the magnetic field as well as the plasma temperature and density.  Using microwave data, \citet{Chen2020b} found that the magnetic field in the loop-top region of the main, brightest part of the flare is about 500 G during the early impulsive phase at 15:54 UT, and \citet{Fleishman2020} found that within a few minutes, the magnetic field had decayed to $\sim$200-300 G.  At an even later time in the gradual phase, 16:28 UT, the magnetic field strength in the main flare arcade had decreased further to 50-200 G, according to an examination of the Stokes profiles of the \ion{Ca}{2} 8542 \AA\ line from the Swedish Solar Telescope \citep{Kuridze2019}. The EOVSA data shown in Figure \ref{eovsa_iris.fig} indicates that the magnetic field in our region of interest (in the loop top region to the south of the main flare arcade) is on the order of 100 G at 15:57 UT, and given the studies above, the field strength has probably decreased somewhat by the time the Doppler shifts are observed in {\it IRIS} after $\sim$16:05 UT.  Therefore, we estimate the magnetic field strength at 16:05-16:15 UT to be 50-100 G.  Using this magnetic field strength to calculate magnetic pressure and our estimates of the temperature and density in this region to calculate the thermal pressure gives a $\beta$ of 0.2--0.8. We compare our observation with the \citet{Takasao2016} model by assuming that the plasma beta in the loop-top region is similar to that in the reconnection inflow region.  In \citet{Takasao2016}, models with values of $\beta$ in the range of 0.2--0.8 exhibit periods of a few hundred seconds  (their Figure 8).  The maximum velocities for these values of $\beta$ in the model are on the order of 100--200 km s$^{-1}$, which is somewhat faster than the velocities we observe with {\it IRIS}. We note, however, that the model in \citet{Takasao2016} is two dimensional, and the averaging effects of integrating over the line of sight could contribute to lowering the observed velocities.

The \citet{Takasao2016} model was built to explain quasi-periodic fast-mode (QPF) magnetosonic waves propagating away from the flare site. In this event, there are no obvious funnel-shaped, QPF wave trains off-limb in the immediate neighborhood of the flare as have been observed in previous events \citep[e. g.][]{LiuW2011}.  However, a study by \citet{LiuW2018} does find some possible QPF signals projected on the disk, and captured in time slices originating from the flare (see their Figure 4b).  These signals have periods of $\sim$4 min, similar to the periods we observe in the Doppler shift in the {\it IRIS} data, so it is possible that the two phenomena are related.

\section{Conclusions}\label{conclusions.sec}

We examine the loop-top region in the 10 September 2017 X8.2 eruption and identify oscillating plasma motions via Doppler shift measurements of the {\it IRIS} Fe XXI line.  We find that the oscillations are damped, and their periods are on the order of 400 s.  Using two different methods, we find that the temperature and density in the region where these oscillations are observed are around 14-16 MK and $\sim$2 $\times$10$^{10}$ cm$^{-3}$, respectively.  We use these values to calculate the loop cooling time, and find that flare loops appear in {\it IRIS} SJI 1330 \AA\ images (dominated by the \ion{C}{2} line sensitive to transition region temperature plasma) after the cooling time has passed.  The locations of the oscillations move southward along the flare arcade, coincident with brightenings of flare ribbons seen in the AIA 1700 \AA\ channel.  A few minutes before the oscillations appear in {\it IRIS}, microwave data from EOVSA indicates that non-thermal electrons are present in the location where the oscillations are seen.

The oscillation periods do not appear to correlate with loop length, ruling out a slow acoustic mode \citep[e.g.]{Wang2003}.  Instead, we suggest that these oscillations are the result of loop-top dynamics that occur when the outflow impinges on the reconnected loops below, creating a ``magnetic tuning fork'' as described in \citet{Takasao2016}.  

These observations illustrate the power of high-cadence spectroscopic observations for understanding the dynamics that occur during solar flares.  Future high throughput instruments such as the EUVST instrument on the Japanese Solar-C mission, or the proposed Multi-Slit Solar Explorer (MUSE) will provide pivotal data for understanding the mechanisms that create these dynamics.

\section*{Acknowledgements} 
K.R. acknowledges support from NASA grants NNX17AB82G and 80NSSC18K0732 to SAO. V.P. acknowledges support from NASA grants 80NSSC20K0716 and NNG09FA40C. G.G. acknowledges support from NASA grant NNX15AJ93G to SAO.
B.C. and S.Y. acknowledge support by NSF grants AGS-1654382 and AST-1735405 to NJIT.  G.L. acknowledges grant 80NSSC19K0075. {\it IRIS} is a NASA small explorer mission developed and operated by LMSAL with mission operations executed at NASA Ames Research center and major contributions to downlink communications funded by the Norwegian Space Center (NSC, Norway) through an ESA PRODEX contract.  {\it Hinode} is a Japanese mission developed and launched by ISAS/JAXA, with NAOJ as domestic partner and NASA and STFC (UK) as international partners. It is operated by these agencies in co-operation with ESA and NSC (Norway). EOVSA operation is supported by NSF grant AST-1910354. This work has benefited from the use of NASA's Astrophysics Data System. CHIANTI is a collaborative project involving researchers at the universities of Cambridge (UK), George Mason and Michigan (USA). 

\bibliographystyle{apj}

\end{document}